# Investigating Unipolar Switching in Niobium Oxide Resistive Switches: Correlating Quantized Conductance and Mechanism

Sweety Deswal[a,b], Ashok Kumar[a,b], and Ajeet Kumar[a,b*]

**Affiliations:**

[a]Academy of Scientific and Innovative Research, CSIR-National Physical Laboratory campus, Dr. K. S. Krishnan Marg, New Delhi 110012, India.

[b]CSIR-National Physical Laboratory, Dr. K.S. Krishnan Marg, New Delhi 110012, India.

[*]Correspondence to: kumarajeet@nplindia.org

## Abstract

Memory devices based on resistive switching (RS) have not been fully realised due to lack of understanding of the underlying switching mechanisms. Nature of ion transport responsible for switching and growth of conducting filament in transition metal oxide based RS devices is still in debate. Here, we investigated the mechanism in Niobium oxide based RS devices, which shows unipolar switching with high ON/OFF ratio, good endurance cycles and high retention times. We controlled the boundary conditions between low-conductance insulating and a high-conductance metallic state where conducting filament (CF) can form atomic point contact and exhibit quantized conductance behaviour. Based on the statistics generated from quantized steps data, we demonstrated that the CF is growing atom by atom with the applied voltage sweeps. We also observed stable quantized states, which can be utilized in multistate switching.

==============

The conventional silicon-based 'flash' memory has reached its limitation due to its slow programming speed, poor endurance and relatively high operating voltage.[1,2] As an alternative, memristive systems have attracted extensive research interest due to their potential usage as components for non-volatile memory[2,3] as well as for logic[4,5] and, brain-inspired neuromorphic computing.[6-8] These two-terminal devices have advantages due to their scalability down to atomic level, high-density storage, low-power consumption, and high-speed features.[6,9-11] The memristive devices show insulating properties on macroscopic scale, but reversibly switch into ionic-electronic conductor at nanoscale dimensions due to formation/dissolution of localized conducting filaments. The conductive filaments are believed to be composed of cations[12,13] or oxygen vacancies.[14-18] Despite the promises shown for its applications, greater insight into the mechanism of CF formation and rupture is required to have control over the devices.

Memristive devices can be classified into bipolar[13,19] and unipolar[13,20,21] types. In bipolar devices, the voltage polarities between the transition from high-resistance state (HRS) to low-resistance state (LRS) and that from LRS to HRS are opposite.[13] However, in unipolar switching the SET process as well as the RESET process can be induced by the same bias voltage polarity. Although, bipolar devices have better endurance, power consumption and variability, the unipolar RS memories have several advantages for logic-in-memory applications. The single polarity operation can solve the sneak path problem in crossbar RRAM leading to very high integration density,[22] also it can simplify the peripheral control complexities in large scale integration circuits.[23] One of the first ReRAM storage test chip was made with unipolar devices by SanDisk and Toshiba.[24] The mechanism of the bipolar switching has been explored in much greater details as compared to the unipolar switching.[19,25-28] The unipolar mechanism is associated with the thermochemically induced stiochiometry variations and redox processes.[13,21] And one of the fundamental question which is still being understood is: how do the conducting filaments grow with the applied voltage?

To explore the mechanism of growth of the CF, we investigated Pt/Nb$_2$O$_5$/Al devices, which showed unipolar switching with high ON/OFF ratio, good endurance cycles and high retention times. We controlled the growth of the CF in the devices to form atomic point contact and observed quantization of conductance with integer (n) and half-integer (n+1/2) multiples of quantum of conductance ($G_0 = 2e^2/h$ ~77.4 μS) states during current vs voltage (I-V) measurements. The quantum conductance statistics analysis indicates that the CF is growing in diameter, atom-by-atom, when the voltage is increasing during the SET step. We also inferred that the CF is a mixed composition of metal cations and oxygen vacancies with high metal ratio.

The Pt/Nb$_2$O$_5$/Al based devices were fabricated over Pt/TiO$_2$/SiO$_2$/Si wafer (commercially available). The Niobium oxide thin films (30 nm) were deposited using reactive magnetron sputtering from metallic Nb target using a dc power in an atmosphere of 94% Ar and 6% O$_2$ at room temperature. The base and deposition pressure of the chamber were kept at ~7×10$^{-7}$ mbar and 2.0×10$^{-2}$ mbar, respectively. Top electrode of Al (~ 100 nm) was deposited using thermal evaporation with dimension of a×a μm$^2$ (a= 100, 200, 300, 400, 500) using shadow mask. The electrical dc measurement of the devices was performed at room temperature using two probe system equipped with Agilent B2901A source-meter. The voltage bias was applied to the top Al electrode while the bottom Pt electrode was grounded. A current compliance was applied during each voltage sweep to avoid permanent breakdown of the device.

The Pt/Nb$_2$O$_5$/Al devices showed unipolar resistive switching (Fig. 1a) with SET, from HRS to LRS, in the range of 1.5-1.9V, and RESET back to HRS in the range of 0.4-0.8 V, with 10$^3$ or higher ON/OFF ratios and high retention times in both HRS and LRS (Fig. 1b). Initially, the pristine devices were in HRS with resistance of the order of *giga-Ohm*. The devices were first needed to undergo electroformation at around 3.5–4 V, as shown in the inset of Fig. 1a. After electroformation, upon positive voltage sweep during SET, the devices switched to LRS with a wide range of resistances, ~30 Ω to ~20 kΩ, which were controlled by applying different current compliances ($I_c$). Figure 1a shows three different unipolar resistive switching cycles of the Pt/Nb$_2$O$_5$/Al cell with $I_c$ values 10 mA, 1 mA and 0.1 mA, resulting in LRS resistances in range of 30 Ω, 1 kΩ and 18 kΩ, respectively. Thus, the strength of CF formed could be controlled by varying $I_c$ during SET. Figure 1c shows 60 switching cycles of one such device with $I_c$ = 1 mA, however, the device did not die or degrade for many more cycles.

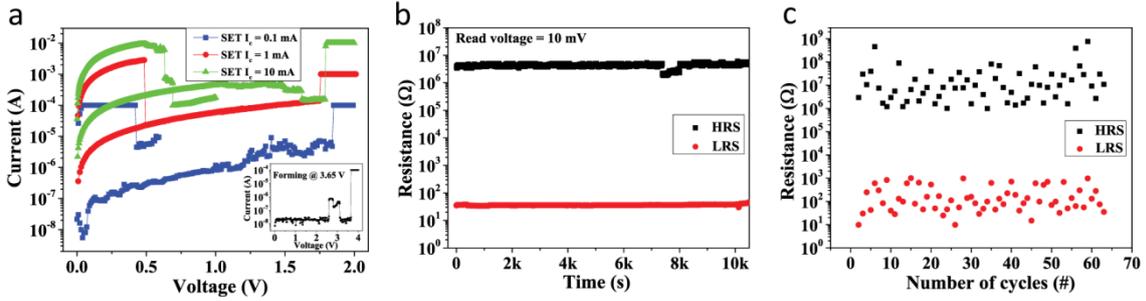

**FIG. 1.** (a) Semi-logarithmic *I-V* plot of Pt/Nb$_2$O$_5$/Al device showing reproducible unipolar resistive switching with SET and RESET. Three different unipolar resistive switching cycles with SET compliance current ($I_c$) values 10 mA, 1 mA and 0.1 mA are shown as green, red and blue I-V curves, respectively. The inset shows electroformation step at ~3.65V with $I_c$ = 0.1 mA. (b) The retention of HRS and LRS for more than 10$^4$ seconds at a read voltage of 10 mV. The LRS is obtained with $I_c$ = 10 mA. (c) Devices switch reproducibly between HRS and LRS and 60 such cycles of switching of a single device is demonstrated. All SET cycles are performed with $I_c$ = 1 mA.

The CFs in the LRS, achieved with SET $I_c$>1 mA, showed *ohmic I-V* characteristics with resistance in the range of 20-50 Ω (supplementary Fig. S1). However, the CFs formed with $I_c$=0.1 mA, showed non-linearity in *I-V* traces with plateaus, which is a non classical behaviour. A few representative curves, having stable step features, are shown in the Fig. 2. The *I-V* traces are shown with G$_0$ on the *y-axis*. The plateaus appeared around the interger or half-interger multiples of G$_0$. After the SET step with $I_c$=0.1 mA, during the RESET of the devices, we observed these stable conductance plateaus due to the formation of an atomic point contact, which is responsible for the quantization in the energy levels of the filament. Similar observations of atomic point contact filament displaying quantization of conductance have been reported for various bipolar[29-31] and unipolar[32-34] devices. The quantized states were observed to be stable for hundreds of seconds around some $G_0$ at read voltage 10mV, indicating that the atomic point contacts of CF can remain stable around one particular quantized state (Supplementary Fig. S2). Stability around quantized states shows the potential of these devices for multilevel memory storage.

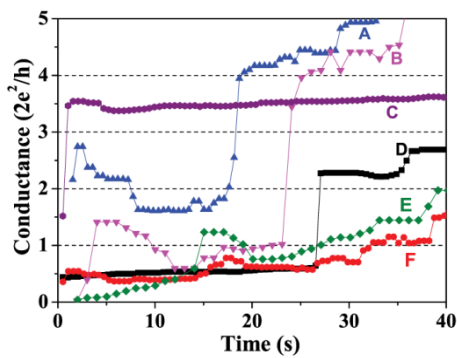

**FIG. 2**. Conducting filaments formed with $I_c$=0.1 mA showed quantization of conductance during *I-V* sweep. The traces (A to F) were stable around some $G_0$ during *I-V* measurements when the voltage was swept for curves A to F in voltage range from 0.0 V to 0.08 V, 0.18 V, 0.38 V, 0.09 V, 0.6 V and 0.38 V, respectively in 40 seconds.

Besides the stable quantized states of LRS in Pt/Nb$_2$O$_5$/Al devices, we observed plateaus, transiently, in *I-V* traces during the SET process. We observed series of quantized conductance plateaus in the SET *I-V* traces, irrespective of any $I_c$ value. Figure 3a shows the evolution of transient quantized states during a SET cycle with $I_c$=0.5 mA. The quantized conductance plateaus again appeared in both integer and half integer multiples of $G_0$. These quantized conductance steps were observed in multiple devices with some of the quantum conductance levels missing on random basis. The histogram of observed quantized conductance steps during 50 SET cycles is plotted in Fig. 3c. We used a total of 8 devices from one fabrication batch to collect all the presented data. We performed statistical analysis of measured conductance quantization, where we used an objective method,[35] which does not make use of any data selection. The data was sorted in bin size of 0.1 $G_0$ and respective numbers were counted, scaled and background noise is subtracted to plot the conductance histogram shown in Fig. 3c. We observed similar quantized conductance step evolution for all devices, irrespective of device size (100 to 500 μm$^2$).

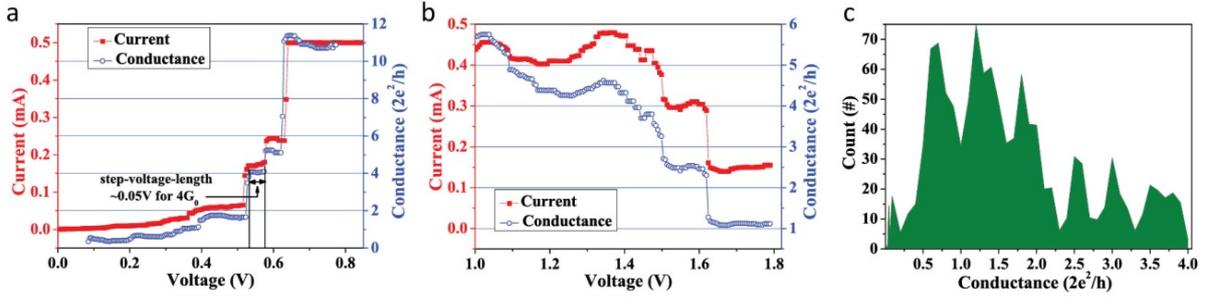

**FIG. 3.** (a) and (b) show quantized conductance plateaus with current on the left scale and conductance on the right scale during SET cycles. The stable current steps were observed resulting in various conductance plateaus. Conductance trace in (a) shows quantized steps at 1, 2, 4, 5, and 7$G_0$. The step-voltage-length of ~0.05V is shown for 4$G_0$. Graph (b) shows $G_0$ values decreasing in steps with increasing voltage bias due to joule heating thermochemical effect causing diffusion of CF atoms, competing with the growth process where CF diameter is growing with increasing $G_0$. This trace is a part of the SET cycle with $I_c$=10 mA presented in Fig. 1a. (c) Histogram of quantum conductance values obtained from *I-V* traces of the devices. The conductance peaks appeared at an equal interval of half-integer multiples of $G_0$.

To understand the growth of the filament on atomic scale, we analyzed step-voltage-lengths, which, we defined as the voltage span for which the plateaus remain stable near one particular $G_0$ before jumping to a lower or higher $G_0$, and were calculated for 0.5$G_0$ to 4$G_0$ in steps of 0.5$G_0$. As an example, step-voltage-length of ~0.05 V for 4$G_0$ is shown in Fig.3a. The range of step-voltage-length, obtained for each $G_0$, was plotted against the corresponding $G_0$ as box chart graph in Fig. 4. We observed that the step-voltage-length is relatively constant, 0.04-0.09 V, for all measured $G_0$ values. We can understand this by considering ballistic carrier transport in one dimensional constriction.[36,37] Electrons flow through a set of discrete one dimensional sub-bands, created due to confinement of charge carriers in one dimensional quantum constriction and each sub-band contribute a conductance unit of $2e^2/h$ or, each conductance transition corresponds to addition of a new sub-band to the channel.[38,39] With assumptions that the transport is ballistic[32] and contacts with electrodes are *ohmic*, the sub-band spacing can be estimated from step-voltage-length at each $G_0$. This relatively constant step-voltage-length suggests that the theoretical increase in the sub-band spacing, with increasing $G_0$, is being compensated by the growing constriction diameter. Due to increased cross section of point contact constriction, we observed decrease in resistance of the filament with subsequent increase in $G_0$ steps, e.g. from ~11.8 k$\Omega$ for 1$G_0$ to 2.5 k$\Omega$ for 5$G_0$. More sub bands are allowed as the radius of conducting filament constriction increases, facilitating conductance jump by one unit of $G_0$ with each additional sub-band. Thus, the constant step-voltage-length with increasing diameter of CF indicates that the CF is forming atom-by-atom.

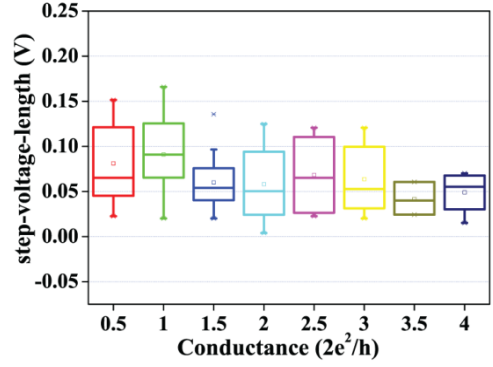

**FIG. 4.** Box chart graph shows step-voltage-length for which the various conductance plateaus remain stable at one $G_0$ before jumping to lower or higher $G_0$. The step-voltage-length is roughly constant (0.04-0.09 V) for each conductance value.

At several instances, during the growth of CF, the $G_0$ decreases in steps with increasing voltage bias. Figure 3b shows a conductance trace which is a part of the SET cycle with $I_c$=10 mA, presented as green curve in Fig. 1a. This may happen due to joule heating thermochemical effect in the unipolar switching causing diffusion of CF atoms, competing with the growth process where CF diameter is growing with increasing $G_0$. However, for the shown SET trace in Fig. 3a, the current is relatively low (0.5 mA) to be able to trigger thermochemical diffusion leading to complete RESET of the device.

Quantization of conductance of the filaments formed in the devices is possible only when the filament is formed with atomic point contact and for this the shape of the filament has to be conical or tapered. The *ohmic I-V* characteristics of the CFs of the LRS also suggest that the composition of the filament is highly rich in metal content. In several reports, the LRS with unipolar devices have been investigated using TEM and have been shown to form metal rich filaments with tapered shape towards anode.[15,27,40,41] When we investigated the resistance of the filament in the temperature range of 293- 353 K, the resistance did not increase with temperature; however, there is no appreciable decrease in the overall resistance of CFs, either (Fig. 5). This suggests that CFs formed in the devices have mixed composition with high metal ratio, which is similar to reported observations.[15,27,40,41]

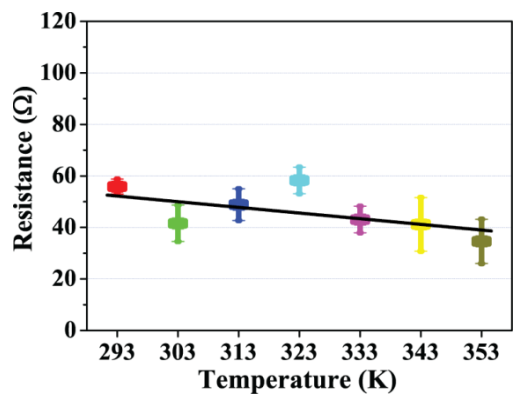

**FIG. 5.** Temperature dependence of LRS from 293 to 353 K shows no appreciable change in the resistance of the conducting filament. The fitting line in black serves as a guide to the eye. The temperature dependent resistance measurement was carried out in ambient condition. Also, in ambient condition, LRS at ~50 Ω shows ±5 Ω variation during $10^4$ second retention cycle.

In summary, we investigated mechanism of conducting filament formation in unipolar Pt/Nb$_2$O$_5$/Al RS cells, which switch with high ON/OFF ratio, good endurance cycles and high retention times. The growth of the filament was controlled using SET compliance current to form atomic point contact, which showed stable quantized conductance steps in *I-V* traces. The steps appeared around integer and half-integer multiples of $G_0$. A step-voltage-length analysis indicated that the filament is growing in diameter, atom-by-atom, with the increase in voltage during *I-V* sweep. The conducting filament appeared to have an admixture of metal cations and oxygen vacancies with high metal-to-oxygen ratio. Also, the devices exhibited stable quantized levels, which is a suitable feature for increasing memory storage density using multilevel cells. Overall, the presented correlation between the atomic point contact structure, conductance and mechanism of formation of conducting filament, provide new insights in the unipolar resistive switching mechanism in oxide-based memristors.

**Acknowledgements:**

Authors would like to thank Dr. Jiji Pulikkotil for valuable discussion. S. D. would like to thank University Grant Commission (UGC) under senior research fellowship (SRF) for financial support and Rahul Kumar and Rupali Malode for assistance in experimental work. This work was financially supported partly by DST INSPIRE Fellowship and partly by CSIR network project AQuaRIUS (PSC0110).

# Supplementary Information

# Investigating Unipolar Switching in Niobium Oxide Resistive Switches: Correlating Quantized Conductance and Mechanism

Sweety Deswal[a,b], Ashok Kumar[a,b], and Ajeet Kumar[a,b*]

[a]Academy of Scientific and Innovative Research, CSIR-National Physical Laboratory campus, Dr. K. S. Krishnan Marg, New Delhi 110012, India.

[b]CSIR-National Physical Laboratory, Dr. K.S. Krishnan Marg, New Delhi 110012, India.

[*]Correspondence to: kumarajeet@nplindia.org

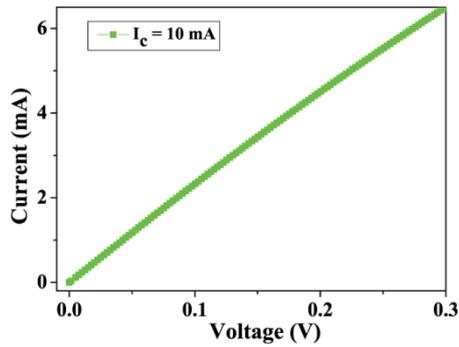

FIG. S1. The figure shows linear *I-V* characteristic in LRS formed with $I_c$=10 mA, suggesting metallic nature of the Conducting Filament.

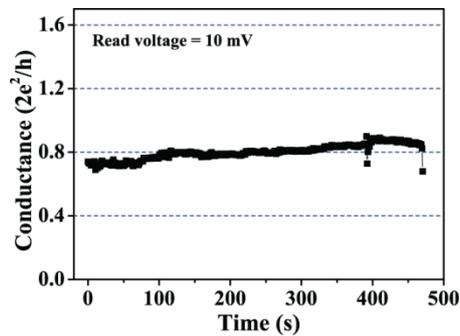

FIG. S2. The figure shows that the quantized conductance level of ~0.8 $G_0$ is stable for around 500 seconds.